\documentclass[11pt,A4paper]{article}
\pdfoutput=1
\usepackage{jheppub}
\usepackage{amsmath}
\usepackage{amssymb}
\usepackage{hyperref}
\usepackage{graphicx}

\def \beq{\begin{equation}}
\def \eeq{\end{equation}}
\def \bea{\begin{eqnarray}}
\def \eea{\end{eqnarray}}

\def \MPlanck{M_{\rm P}}

\def \nn{\nonumber}

\newcommand{\MeV}{\hbox{\,MeV}}
\newcommand{\GeV}{\hbox{\,GeV}}
\newcommand{\TeV}{\hbox{\,TeV}}

\title{Dark Radiation in Anisotropic LARGE Volume Compactifications}

\author{Stephen Angus}

\affiliation{Rudolph Peierls Centre for Theoretical Physics, University of Oxford, \\
1 Keble Road, Oxford, OX1 3NP, United Kingdom}

\emailAdd{Stephen.Angus@physics.ox.ac.uk}

\abstract{ 
Dark radiation is a compelling extension to $\Lambda$CDM: current experimental results hint at $\Delta N_{\rm eff} \gtrsim 0.5$, which is increased to $\Delta N_{\rm eff} \simeq 1$ if the recent BICEP2 results are~included.  In recent years dark radiation has been considered in the context of string theory models such as the LARGE Volume Scenario of type IIB string theory, forging a link between present-day cosmological observations and models of physics at the Planck scale.
In this paper I consider an 
extension of the LARGE Volume Scenario in which the bulk volume is stabilised by two moduli instead of one.  Consequently, the lightest modulus no longer corresponds to the compactification volume but instead to a transverse direction in the bulk geometry.  I focus on scenarios in which sequestering of soft masses is achieved by localising the Standard Model on D3 branes at a singularity.
The fraction of dark radiation produced in such models vastly exceeds experimental bounds, ruling out the sequestered LARGE Volume Scenario with two bulk moduli as a model of the early Universe.}

\begin{document}
\maketitle
\flushbottom

\section{Introduction}
\label{sec:intro}
In recent years there has been speculation about the possible existence of an additional relativistic matter component in the energy density of the Universe.  This so-called dark radiation is motivated both theoretically and phenomenologically.  In UV-complete quantum gravity frameworks such as string theory, the existence of light axion-like particles (ALPs) is commonplace --- string compactifications typically produce hundreds of moduli, with associated axions\footnote{Hereafter we make liberal use of the term ``axion'' to refer to axion-like particles in string compactifications.} that are massless at the perturbative level due to shift symmetries.  Meanwhile, the fact that dark matter is a crucial ingredient in the Standard $\Lambda$CDM Cosmological model implores us to ask: if dark matter, then why not dark radiation?  The number of relativistic particle species is not protected by any symmetry, therefore there is no reason to assume \emph{a~priori} that the present-day radiation content of the Universe must consist of only photons and neutrinos.

Dark radiation is conventionally described in terms of an ``excess effective number of neutrino species,''
\beq
\rho_{\rm DR} = \frac{7}{8}\left(\frac{4}{11}\right)^{4/3}\!\rho_\gamma\,\Delta N_{\rm eff} \, ,
\eeq
where $\Delta N_{\rm eff} = N_{\rm eff} - 3.046$.
There are mounting experimental hints for dark radiation.  Assuming a tensor-to-scalar ratio $r = 0$, a combination of recent CMB observations by Planck \cite{13035076}, high-$l$ data from SPT \cite{12126267} and ACT \cite{13010824}, WMAP 9-year polarisation data \cite{12125226}, BAO measurements, and the value of $H_0$ observed by the Hubble Space Telescope \cite{11032976}, suggests $N_{\rm eff} = 3.52^{+0.48}_{-0.45}$ at 95\% c.l..  Meanwhile, independent constraints from Big Bang Nucleosynthesis give $N_{\rm eff} = 3.50 \pm 0.20$ \cite{13083240}.  The case for dark radiation is further enhanced if one incorporates the recent discovery of primordial B-modes by BICEP2 \cite{BICEP2}: using a $\Lambda$CDM+$r$ model with $r = 0.2^{+0.07}_{-0.05}$, the authors of \cite{14034852} find a preference for dark radiation, with $N_{\rm eff} = 4.00 \pm 0.41$ [Planck+WP+BICEP2] at 68\% c.l.; meanwhile, other similar studies \cite{Zhang:2014dxk,Dvorkin:2014lea} find \hbox{$N_{\rm eff} \sim 3.86 \pm 0.25$} and \hbox{$N_{\rm eff} \sim 3.95 \pm 0.33$} at 68\% c.l., respectively.  Together, these results provide compelling hints for the possible existence of extra relativistic species. 

One of the key motivations for studying dark radiation is that it provides a means of testing models of physics at the Planck scale, such as string theory models.  During inflation, the moduli of string compactifications are displaced from their final VEVs, such that when inflation ends they begin to oscillate about their global minimum.  Since the moduli behave as non-relativistic matter, they eventually come to dominate the energy density of the Universe.  The subsequent reheating of the visible Universe and production of hidden particle species is thus determined by the decay modes of moduli.  

In general, moduli have Planck-suppressed decay rates that scale as their mass cubed,
\beq
\Gamma_\Phi \sim \frac{m_\Phi^3}{\MPlanck^2} \, .
\eeq
Therefore the lightest modulus is the longest-lived, and since radiation redshifts as $a^{-4}$ whereas non-relativistic matter evolves as $a^{-3}$, any radiation produced by early decays will have redshifted away by the time the lightest modulus decays.  Hence reheating is driven solely by the decays of the lightest modulus to the visible sector.  Furthermore, this implies that the lightest modulus is also dominantly responsible for dark radiation production. 

One phenomenologically appealing string theory model is the LARGE Volume Scenario (LVS) of type IIB string theory \cite{LVS,0505076,08051029}.  In the most basic realisation of this scenario, the overall compactification volume $\mathcal{V}$ is determined by a single bulk cycle, while additional smaller blow-up cycles can support the visible sector, non-perturbative effects, and additional hidden sectors.  The bulk volume is controlled by a K\"{a}hler modulus known as the volume modulus, which is stabilised at an exponentially large size due to a combination of $\alpha'$-corrections and non-perturbative effects.  Consequently, this modulus is hierarchically lighter than all the other moduli, with a mass $m_\mathcal{V} \sim \MPlanck/\mathcal{V}^{3/2}$ (whereas all the other moduli are stabilised around the gravitino mass scale, $m_{3/2} \sim \MPlanck/\mathcal{V}$).

The branching fraction to dark radiation has been studied for this minimal LVS \cite{12083562, 12083563} (see also \cite{14014364}), in which 
the primordial abundance of dark radiation is determined by the decays of the volume modulus to visible- and hidden-sector particles.  It turns out that there is one dominant visible-sector decay mode: \hbox{$\Phi \to H_u H_d$} via a Giudice-Masiero term with $\mathcal{O}(1)$ dimensionless coupling $Z$ \cite{Giudice:1988yz}.\footnote{This is a dimension-5 operator, so the overall coupling is $Z/\MPlanck$ times a numerical factor.}  The branching fraction to dark radiation can thus be computed: for the case of a shift symmetry in the Higgs sector, which implies $Z = 1$ at the string scale \cite{12042551}, one finds a lower bound of $\Delta N_{\rm eff} \gtrsim 1.4$.  This is in tension with $\Delta N_{\rm eff} \simeq 0.5$ even after loop effects are taken into account \cite{13054128} --- this is an example of the ``moduli-induced axion problem'' \cite{13047987}, which is the statement that string models generically produce too much dark radiation via decays to axion-like particles.  However, this tension is relaxed significantly if the BICEP2 results are included in the analysis: a value of $N_{\rm eff} = 4.00 \pm 0.41$ \cite{14034852} is compatible with the minimal LVS, with disagreement at only the $1\sigma$ level.

It is worthwhile to investigate whether or not extended models can yield a value of $\Delta N_{\rm eff}$ that is compatible with observations.  A simple extension of LVS is the scenario in which the bulk volume is controlled by two K\"{a}hler moduli instead of one \cite{08080691,11052107,11106182,12036655,12081160}.  One linear combination of these two moduli is the volume modulus, while a transverse flat direction remains unstabilised in the tree-level potential.

Such a setup has a fibration structure and may lead to anisotropic modulus stabilisation.  However, as I will discuss in section \ref{sec:anisotropic}, anisotropy is \emph{not} an essential requirement, and the conclusions of this paper apply to all fibred models with a particular sequestered structure. 
In fact, the crucial feature of these compactifications most relevant to our purposes is that the volume modulus is no longer the lightest modulus: the post-inflationary decays to visible and hidden radiation are instead controlled by the modulus parametrising the transverse direction.  Hence this extension has non-trivial consequences for post-inflationary physics, and one might imagine that the above constraints on dark radiation could thus be avoided.  The purpose of this paper is to analyse such a scenario and determine how the branching fraction to dark radiation is modified. 

The structure of this paper is as follows.  In section \ref{sec:anisotropic} I describe and justify a two-modulus compactification scheme, for which I compute the decay modes, and deduce the consequences for $\Delta N_{\rm eff}$, in section \ref{sec:decays}.  In section \ref{sec:conclusions} I conclude and discuss a possible alternative scenario.

\section{Fibred compactifications}
\label{sec:anisotropic}
Here we give an overview of some key features of fibred LVS models. 
First of all, the compactification volume $\mathcal{V}$ takes the form\footnote{Examples of such compactifications are K3 or $T^4$ fibrations over a $\mathbb{P}^1$ base.}
\beq \label{eq:vol}
\mathcal{V} = \alpha\sqrt{\tau_1}\tau_2 - \sum_{i=3}^{h_{1,1}^+}\beta_i\tau_i^{3/2} \, ,
\eeq
where $\tau_1$ and $\tau_2$ are the K\"{a}hler moduli that determine the bulk extra-dimensional volume (\hbox{$\tau_1$ corresponds} to the fibre volume while the combination $t_1 \sim \tau_2/\sqrt{\tau_1}$ gives the volume of the base), and the remaining $\tau_i$ describe blow-up cycles (``holes'') in the geometry.  Such a model will also have $h_{1,1}^+$ axions $a_i$, so we can define complexified K\"{a}hler moduli, $T_i \equiv \tau_i + ia_i$. In the following section we will neglect all moduli except for $T_1$ and $T_2$, since it turns out that we are focussing on energy scales at which all the other moduli (including complex structure moduli and the axio-dilaton) can be integrated out.


We now wish to stabilise these moduli.  In particular, by considering the case of a Euclidean D3 (ED3) brane wrapping one of the blow-up moduli (say $\tau_3$) and stacks of D7s wrapping $\tau_1$ and $\tau_2$,\footnote{This is the ``small hierarchy'' scenario of \cite{11052107}.  We will not consider the large-hierarchy case at present, since stabilising $\tau_1$ and simultaneously avoiding the CMP requires some additional tuning ($\tau_1 \sim p$ \cite{12116927} while $m_\Omega \sim \MPlanck/\mathcal{V}^{(3+p)/2}$ \cite{11052107}).
However, see section \ref{sec:conclusions} for an interesting realisation of the large-hierarchy scenario.} one obtains a non-perturbative superpotential of the form
\beq
W = W_0 + Ae^{-aT_3} \, .
\eeq
Here $W_0$ is the tree-level superpotential, which is independent of the K\"{a}hler moduli, and the second term is a non-perturbative correction due to instanton effects.  This scenario leads to stabilisation of $\mathcal{V} \sim \sqrt{\tau_1}\tau_2$ at an exponentially large value, while the flat transverse direction is lifted by string loop corrections, which arise owing to the D7 stacks on the bulk cycles $\tau_1$ and $\tau_2$.

This setup typically leads to stabilisation of the bulk moduli at exponentially large values, with $\tau_2 \sim \tau_1$ \cite{08080691,11052107}. However, the precise relationship depends on the coefficients of the string loop corrections, which in turn depend on the complex structure moduli and other details of the compactification.  Given the exponentially large bulk volume, these moduli may be separated by several orders of magnitude. Hence such a setup can easily lead to an anisotropic geometry, with two of the extra dimensions some orders of magnitude larger or smaller than the other four.  
Nevertheless, we re-emphasise that 
the conclusions of this paper are not restricted to the anisotropic limit and in fact apply when $\tau_1 \sim \tau_2$.

When constructing a realistic model we must bear in mind low-energy phenomenological constraints.  In particular, we would like to ensure that soft terms in the visible sector are realised at a scale sufficiently suppressed relative to the masses of all moduli.  If this were not the case, requiring TeV-scale superpartners would bring the moduli down to scales $m_\Phi~\lesssim~30\TeV$.  Such low moduli masses encounter the Cosmological Moduli Problem (CMP), in which they dominate the energy density of the Universe at a scale low enough to spoil the successful BBN predictions \cite{CMP1, CMP2, CMP3}.  This problem can be avoided if the Standard Model is realised on D3 branes at a singularity, leading to a sequestering of the scale at which soft masses appear, as illustrated below. 

In summary, the conclusions we draw from this analysis apply to all LVS models in which:
\begin{itemize}
\item the volume takes the form (\ref{eq:vol});
\item the low-energy spectrum of moduli contains $\tau_1$, $\tau_2$, and their axions;
\item the modulus corresponding to the transverse direction is stabilised at a scale parametrically lighter than the volume modulus;
\item the Standard Model is located on D3 branes at a singularity.
\end{itemize}
This includes most fibred models --- exceptions include scenarios in which the Standard Model is realised via D7 branes on the fibre cycle, or in which the fibre modulus $\tau_1$ is stabilised by D-term constraints (for example, as in the scenarios considered in \cite{14036810}).  

\subsection{Mass hierarchy}
\label{sec:masses}
In this particular 
realisation of LVS, a distinctive hierarchy of mass scales is generated \cite{0505076, 11052107, 10055076}.  After diagonalising in terms of mass eigenstates, we find that
\begin{align} \label{eq:masses}
m_{\tau_i} \sim m_{a_i} &\sim \frac{\MPlanck\ln\mathcal{V}}{\mathcal{V}} \, \;\;\;\; (i \neq 1, 2) \, , \nn \\
m_{3/2} \sim  m_S \sim m_U &\sim \frac{\MPlanck}{\mathcal{V}} \, , \nn \\
m_{\mathcal{V}} &\sim \frac{\MPlanck}{\mathcal{V}^{3/2}} \, , \nn \\
m_{\Omega} &\sim \frac{\MPlanck}{\mathcal{V}^{3/2}\tau_1^{1/4}} \, , \nn \\
m_{a_1} \sim m_{a_2} &\simeq 0 \, .
\end{align}
Here $\Omega$ is the direction transverse to the volume; in the limit $\tau_1 \to \tau_2$, $m_{\Omega} \to \MPlanck/\mathcal{V}^{5/3}$.  
For scenarios in which the Standard Model is located on branes at a singularity \cite{09063297}, soft masses are expected to appear at a scale $M_{\rm soft}~\sim~\MPlanck/\mathcal{V}^2$; for TeV-scale superpartners this implies a volume $\mathcal{V}~\sim 5~\times~10^7\GeV$, so $m_\mathcal{V}~\sim~3~\times~10^6\GeV$ and $m_\Omega~\sim 10^5\GeV$ (for $\tau_1 \sim \tau_2$).\footnote{If there is no sequestering, the soft masses instead arise at a scale $m_{3/2}~\sim~\MPlanck/\mathcal{V}$, which is incompatible with reheating via the decay of the lightest modulus (which also suffers from the CMP).}

Let us briefly comment on the values that the VEV of $\tau_1$ can take.  On one hand we require $\tau_1 \gg 1$ --- if this were not the case, (\ref{eq:masses}) indicates that the volume and transverse moduli would have comparable masses, so we would not be able to neglect the volume modulus interactions.  Furthermore, if $\tau_1 \lesssim \mathcal{O}(1)$ in string units the perturbative $\alpha'$ expansion is no longer trustable.  On the other hand, we cannot take $\tau_1$ too large as this will reduce $m_\Omega \lesssim 30\TeV$, so we encounter the CMP again --- for TeV-scale SUSY this corresponds to $\tau_1 \gtrsim 10^9$.  We avoid all of the above problems by taking ``natural'' values, $\tau_1 \sim \tau_2 \sim 10^4 \text{--} 10^6$.

\section{Leading decay modes}
\label{sec:decays}
We now turn to the computation of the leading decay modes of the lightest modulus $\Omega$.  In this analysis we focus on the branching fractions to the bulk axions and to visible matter, neglecting other possible hidden-sector channels (some of which, such as additional closed-string axions, could also contribute to dark radiation).

\subsection{Dark radiation}
\label{sec:dr}
The decay to axions can be computed from the K\"{a}hler potential for the bulk K\"{a}hler moduli, which can be expressed as
\beq
K = -\ln(T_1 + \bar{T}_1) - 2\ln(T_2 + \bar{T}_2) \, .
\eeq
This is simply the expansion of the usual 4-dimensional $\mathcal{N} = 1$ supergravity K\"{a}hler potential for the K\"{a}hler moduli, $K = -2\ln\mathcal{V}$, in the 
fibred scenario (up to an irrelevant constant term).  From this, we generate un-normalised kinetic terms of the form
\beq
\mathcal{L} \supset \frac{1}{4\tau_1^2}\left(\partial_\mu\tau_1\partial^\mu\tau_1 + \partial_\mu a_1 \partial^\mu a_1\right)
+ \frac{1}{2\tau_2^2}\left(\partial_\mu\tau_2\partial^\mu\tau_2 + \partial_\mu a_2 \partial^\mu a_2\right) \, .
\eeq

We can canonically normalise the moduli with the reparametrisation
\beq
\Phi_1 = \frac{1}{\sqrt{2}}\ln\tau_1 \, , \quad \Phi_2 = \ln\tau_2 \, ,
\eeq
which once expanded out gives
\begin{align}
\mathcal{L} \supset &\frac{1}{2}\partial_\mu\Phi_1\partial^\mu\Phi_1 + \frac{1}{2}\partial_\mu a_1\partial^\mu a_1
+ \frac{1}{2}\partial_\mu\Phi_2\partial^\mu\Phi_2 + \frac{1}{2}\partial_\mu a_2\partial^\mu a_2 \nn \\
&- \sqrt{2}\Phi_1\partial_\mu a_1\partial^\mu a_1 - \Phi_2\partial_\mu a_2 \partial^\mu a_2 \, .
\end{align}
The second line of this expression contains the relevant interactions for decays of the $\Phi$ moduli to axions, which according to (\ref{eq:masses}) are massless and therefore constitute dark radiation.

To extract the relevant physics, we must rotate $\Phi_1$ and $\Phi_2$ into their mass eigenbasis \cite{10054840},
\beq
\Phi_\mathcal{V} \equiv \sqrt{\frac{2}{3}}\Phi_2 + \sqrt{\frac{1}{3}}\Phi_1 \, , \qquad
\Phi_\Omega \equiv \sqrt{\frac{1}{3}}\Phi_2 - \sqrt{\frac{2}{3}}\Phi_1 \, , \,
\eeq
where $\Phi_\mathcal{V}$ is the volume modulus and $\Phi_\Omega$ is the transverse flat direction.
Since $\Phi_\Omega$ is the lightest modulus its decays will dominate, and hence the relevant interaction for decays into dark radiation is
\beq
\mathcal{L}_{\Omega \to aa} =
\frac{1}{\sqrt{3}\MPlanck}\Phi_\Omega\left(2\partial_\mu a_1\partial^\mu a_1 - \partial_\mu a_2\partial^\mu a_2\right) \, .
\eeq
This yields a total decay rate to axions of
\beq
\Gamma_{\Omega \to aa} = \frac{5}{96\pi}\frac{m_\Omega^3}{\MPlanck^2} \, ,
\eeq
which is a factor $5/2$ larger than in the minimal LVS.

\subsection{Visible sector}
\label{sec:vis}

In the minimal LVS model \cite{12083562} the leading decay mode is to Higgs bosons via the Giudice-Masiero term \cite{Giudice:1988yz},
\beq \label{eq:higgsmin}
\mathcal{L} \supset  \frac{1}{\sqrt{6} \MPlanck} \, \Big [ Z  \hspace{0.25mm} H_u H_d  \hspace{0.5mm} \Box \hspace{0.25mm}  \Phi + {\rm h.c.} \Big  ] \,.
\eeq
Let us see how this is modified in the 
present case.  We can compute the relevant Lagrangian from a K\"{a}hler potential of the form
\beq
K = -\ln(T_1 + \bar{T}_1) - 2\ln(T_2 + \bar{T}_2) + \left\{\frac{H_\text{u}\bar{H}_\text{u} + H_d\bar{H}_d + (ZH_uH_d + \rm{h.c.})}{\big(T_1 + \bar{T}_1\big)^{1/3}\big(T_2 + \bar{T}_2\big)^{2/3}}\right\} \, .
\eeq
This expression contains all the relevant physics: in particular, it incorporates the appropriate scaling of the K\"{a}hler matter metric with $\mathcal{V}^{-2/3}$ \cite{12116927}.

Extracting the leading terms, one finds the interaction Lagrangian
\begin{align} \label{eq:decaytohiggs}
\mathcal{L} \supset 
&- \frac{1}{\sqrt{6}\MPlanck}\left(\sqrt{\frac{1}{3}}\Phi_1 + \sqrt{\frac{2}{3}}\Phi_2\right)\left[H_u\Box\bar{H}_u + \bar{H}_u\Box H_u
+ H_d\Box\bar{H}_d + \bar{H}_d\Box H_d\right] \nn \\
&- \frac{1}{\sqrt{6}\MPlanck}\left(ZH_uH_d + \rm{h.c.}\right)\left(\sqrt{\frac{1}{3}}\Box\Phi_1 + \sqrt{\frac{2}{3}}\Box\Phi_2\right) \, .
\end{align}
The second line contains the dominant interactions, as the $\Box\Phi$ terms induce a scaling with $m_\Phi^2$, which from (\ref{eq:masses}) is a factor $\mathcal{V}^{1/2} \sim 10^3$ larger than $m_H^2$.  The dominant terms have the same structure as (\ref{eq:higgsmin}), however, note that the moduli always appear in the combination $\sqrt{1/3}\Phi_1 + \sqrt{2/3}\Phi_2 \equiv \Phi_\mathcal{V}$.  In particular, the Lagrangian is independent of the lightest modulus $\Phi_\Omega$, so this decay mode is suppressed  at tree-level.\footnote{The volume $\mathcal{V}$ does have a subleading dependence on $\Phi_\Omega$, which has been computed in \cite{12024580}.  However, the resulting term is suppressed by $\mathcal{V}^{-1/3} \sim 3 \times 10^{-3}$, so its contribution can be neglected.}

Let us consider other possible decay modes of $\Phi_\Omega$.  Chiral matter scalars also interact only with $\Phi_\mathcal{V}$ at tree-level (the relevant Lagrangian has the same form as the first line of (\ref{eq:decaytohiggs})), while interactions with fermions are chirality-suppressed because they will always contain the Dirac operator $\bar{\chi}\bar{\sigma}^\mu\partial_\mu\chi$, which vanishes on-shell.  Furthermore, since the Standard Model is localised on a blow-up cycle, interactions of gauge bosons with the bulk moduli $\Phi_\mathcal{V}$ and $\Phi_\Omega$ will be volume-suppressed, so decay via gauge bosons also takes place only at loop level.  Finally, if there are additional vector-like matter states, we expect their tree-level couplings to moduli to be of the same form as (\ref{eq:decaytohiggs}) and hence also independent of $\Phi_\Omega$.  We conclude that all visible-sector decay modes must arise only at loop level, so decays to dark radiation (and possibly other hidden-sector particles) are the dominant processes.


\subsection{Prediction for the excess effective number of neutrino species}
\label{sec:dneff}
We now provide an estimate for $\Delta N_{\rm eff}$ based on these conclusions.  Assuming the relevant loop-level decay rates have the form
\beq
\Gamma_{\rm 1-loop} \sim \frac{1}{16\pi}\left(\frac{\alpha_{\rm SM}}{4\pi}\right)^2\frac{m_\Omega^3}{\MPlanck^2} \, ,
\eeq
where $\alpha_{\rm SM}$ represents visible-sector couplings, the ratio of hidden to visible branching ratios is
\beq
\kappa \equiv \frac{\rm Br(hidden)}{\rm Br(visible)} \sim \frac{5}{6}\left(\frac{4\pi}{\alpha_{\rm SM}}\right)^2 \sim 10^{4} \, .
\eeq
Since $\Delta N_{\rm eff} \gtrsim 3\kappa$ \cite{12083562,12083563}, this implies $\Delta N_{\rm eff} \gtrsim 3\times 10^4$, which completely rules out the 
fibred scenario with D3 branes at a singularity as a realistic model of dark radiation.

\section{Discussion and conclusions}
\label{sec:conclusions}
In this paper we have considered dark radiation in a simple fibred extension of the minimal LARGE Volume Scenario. 
We have focussed on 
models in which the visible sector is located on D3 branes at a singularity, which sequesters the soft terms down to order $\MPlanck/\mathcal{V}^2$.  For a compactification volume $\mathcal{V} \sim 5 \times 10^7$ the lightest modulus obtains a mass at $m_\Omega \sim 10^5\GeV$, sufficiently heavy to avoid the Cosmological Moduli Problem.  We have computed the ratio of branching fractions to hidden- and visible-sector final states and found that this scenario is killed by an excess of dark radiation, with $\Delta N_{\rm eff} \gtrsim 3\times 10^4$.  We must therefore turn to other scenarios in order to avoid overproduction of axion-like particles. 


Let us consider one such alternative scenario: a fibred LVS model in the anisotropic limit, $\tau_1 \ll \tau_2$.  One way to introduce such an anisotropy is to demand an appropriate tuning of parameters in the superpotential.  A potentially more robust approach is to examine a concrete scenario in which an anisotropy is generated: an example of such a model is considered in \cite{12116927}.  In this model, the exponentially large volume is realised via stacks of D7 branes on a blow-up cycle $\tau_3$, leading to gaugino condensation on that cycle.  If the D7s are arranged such that there are two separate gauge groups on the cycle that are allowed to condense independently, one obtains a racetrack superpotential of the form
\begin{equation}
W = W_0 + Ae^{-aT_3} - Be^{-bT_3} \, .
\end{equation}

An important new ingredient in this construction is the presence of a Euclidean D3 brane that wraps the fibre cycle $\tau_1$.  It has been argued in \cite{11052107} that for an appropriate fibration (such as a K3 or $T^4$) 
it is in principle possible to suppress the usual instanton contribution to the superpotential from this cycle.  However, the ED3 brane may instead give instanton-like corrections to the gauge kinetic function of a different cycle --- in this case the blow-up cycle $\tau_3$ --- which in turn generates poly-instanton corrections to the superpotential,
\begin{equation}
W = W_0 + Ae^{-a(T_3 + C_1e^{-2\pi T_1})} - Be^{-b(T_3 + C_2e^{-2\pi T_1})} \, .
\end{equation}
This setup leads to anisotropic modulus stabilisation, with $\tau_2 \gg \tau_1 \sim \tau_3$.
Finally, if visible matter is then realised on D7 branes wrapping $\tau_1$ it is possible to achieve a hierarchy between generations of soft scalar masses, with the lightest states separated by a factor $\mathcal{V}^{-1}$ from the rest \cite{12116927}.\footnote{Note that this genarational hierarchy applies only to scalar masses; gaugino masses, on the other hand, are stabilised around the gravitino mass scale, $M_a \sim m_{3/2} \sim M_{\rm P}/\mathcal{V}$.}  This separation of scales allows for the possibility of natural supersymmetry, which could explain the lack of observations of supersymmetric particles at the LHC \cite{0602096}.

An important consequence of such a scenario is that the K\"{a}hler matter metric would no longer depend only on the bulk volume $\mathcal{V}$ but also on the transverse direction $\Omega$, so a tree-level coupling to Higgs bosons would be restored.\footnote{Note however that, as argued below, 
the suppression of the Higgs bilinear is at most $\mathcal{V}^{-1}$ times a loop factor.  Hence the only kinematically viable decay to the Higgs sector would be to the light $125\GeV$ boson observed at the LHC.}  Furthermore, since the Standard Model would now be located on the bulk cycle $\tau_1$, a tree-level coupling to gauge bosons would be generated via the gauge kinetic function, which depends on $T_1$ and hence on the lightest modulus $\Phi_\Omega$.  This further enhances the branching fraction to the visible sector.  The predictions for axionic dark radiation in the anisotropic scenario, with visible matter on D7s wrapping $\tau_1$, have been worked out in \cite{14036810} --- they allow $\Delta N_\text{eff} \simeq 0.6$ for $Z = 1$ at the string scale (in the case where the reheating temperature is sufficiently high that all Standard Model degrees of freedom are thermalised after reheating).

It should be noted that there are a number of technical issues with this particular construction, which would need to be addressed in a fully consistent model.  First of all, it is unclear if poly-instantons actually exist in type IIB string theory.  Second, if such corrections do exist, their origin via the ED3 on the fibre cycle must be reconciled with the additional presence of the visible-sector D7s on the same cycle.  These D7s are likely to generate their own non-perturbative corrections due to gaugino condensation, overwhelming the poly-instantons necessary to generate the anisotropy in the first place.

Third, even if such a construction could be realised it is not clear that it would be stable under loop corrections due to RG running.  Such corrections are likely to reduce the hierarchy between generations of visible-sector scalars to no more than a loop factor.  
Hence it does not immediately appear possible to realise TeV-scale supersymmetry in such a scenario, as to do so would require $\mathcal{V} \sim \mathcal{O}(10^{14})$ and push $m_\mathcal{V}$ down to $\MeV$ scales, leading once again to the Cosmological Moduli Problem. 
Therefore, this scenario appears to make sense only in the context of high-scale supersymmetry, with soft terms \hbox{$M_\text{soft} \gtrsim 10^8\GeV$} --- in particular, natural supersymmetry would require that the soft-term hierarchies discussed above are rendered sufficiently stable under quantum corrections 
that a significant residual hierarchy is maintained.  Overall, given the technical obstacles to realising this version of anisotropic modulus stabilisation, the prospect of simultaneously obtaining natural supersymmetry \emph{and} an acceptable yield of dark radiation in such a scenario, while very appealing, remains somewhat far-fetched.


Finally, we remark that if the recent claimed discovery of primordial B-modes by BICEP2 \cite{BICEP2} holds up, estimates for the effective number of neutrino species could increase $N_{\rm eff}$ to values in the region of $(3.8 - 4)\pm 0.4$ at \hbox{68\% c.l. \cite{14034852,Zhang:2014dxk,Dvorkin:2014lea}.}  Such values are significantly closer to the $\Delta N_{\rm eff} \gtrsim 1.4$ prediction of the minimal LVS, so ultimately the minimal scenario may be saved after all.

\emph{NOTE: This paper was submitted simultaneously to the related work \cite{14036810}.}

\acknowledgments{I would like to thank Lukas Witkowski, Joseph Conlon, Michele Cicoli and Sven Krippendorf for helpful discussions, comments and feedback.  SA is funded by an STFC studentship.}

\end{document}